\documentclass{iau}
\usepackage{graphicx}

\title{The Virgo Cluster}
\author{Jeremy Mould}
\affiliation{$^1$Centre for Astrophysics and Supercomputing, Swinburne University, Hawthorn 3122, Australia\\
[\affilskip]
$^2$ARC Centre of Excellence for All-sky Astrophysics (CAASTRO)\\ email: {\tt jmould@swin.edu.au}\\

}

\pubyear{2012}
\volume{289}  
\pagerange{119--126}
\jname{Advancing the physics of cosmic distances}
\editors{R. de Grijs \& G. Bono, eds.}
\begin{document}

\maketitle

\begin{abstract}
In the era of precision cosmology the Virgo cluster takes on a new role in the cosmic
distance scale. Its traditional role of testing the consistency of secondary distance indicators
is replaced by an ensemble of distance measurements within the Local Supercluster united by
a velocity field model obtained from redshift survey based reconstruction. WMAP leads us to see the Hubble Constant
as one of six parameters in a standard model of cosmology with considerable covariance between parameters.
Independent experiments, such as WMAP and the HST Key Project (and their successors) constrain these
parameters.

\end{abstract}

\firstsection 
\section{Introduction}
At the height of the 1980s distance scale controversy the Virgo cluster had iconic status. 
Now, following the Las Campanas Virgo Cluster survey (Binggeli, Sandage, \& Tammann 1985) and x-ray imaging, we see Virgo as a concentration of matter to be mapped. 
Its place has been taken by an ensemble of distance measurements within z $<$ 0.01, linked by a velocity field model developed from the 2MASS redshift survey.
 Some expansion rate myths are discussed in this review, and how to see past them. 
We should be aiming to measure H$_0$ to 1\% accuracy. 

\section{Virgo galaxies' distances}
Figure 1 is prepared from the NED distance database for Virgo. Similar results have been obtained by
Mei et al (2007) using the Surface Brightness Fluctuations distance indicator. A map of the cluster in projection is furnished by 
Bohringer \etal~ (1994) from ROSAT.

In a dozen years we have moved from the distance scale controversy illustrated by Table 1 from 
Tammann (1999, IAU Symposium 183) to a cluster mapping perspective.
It is noteworthy that the Tully Fisher relation which celebrates its 35th anniversary this year provided the
 first correct measurement of the distance of Virgo, (Mould, Aaronson, Huchra 1980), 31.0 $\pm$ 0.1 mag.
No morphological type dependence was seen  (Aaronson \& Mould 1983).
One might ask, what about two galaxies with the same  $\Delta V$ and seriously different bulges ?
Semi analytic models shows a Tully Fisher morphological type dependence
(Tonini \etal~ 2012) which can be removed by plotting a new dynamical parameter combining velocity dispersion and
rotational velocity / line width. The reason for the weakness of the TF morphological type dependence is the strong correlation of galaxy mass with bulge velocity dispersion
(Catinella \etal~ 2012).

\pagebreak
\leftline{\bf Table 1: Virgo cluster modulus from various methods [IAUS 183]}
\leftline{(Tammann 1999)}
\begin{tabbing}
Methodsssssssssssss\= (m-M)$_{Virgo}$sss\= Hubble type\kill

Method\> (m-M)$_{Virgo}$\> Hubble type\\
Cepheids\> 31.52$\pm$0.21 \>S\\
SNeIa\> 31.39$\pm$0.17 \> E,S\\
Tully-Fisher\> 31.58$\pm$0.24\> S\\
Globular Clusters\> 31.67$\pm$0.15\> E\\
D$_n~\sigma$\> 31.85$\pm$0.19\> S0,S\\
Novae\> 31.46$\pm$0.40\> E\\
Mean:\> 31.60$\pm$0.08\> ($=>$ 20.9$\pm$0.8 Mpc)\\
\end{tabbing}


\section{Secondary distance indicators in Virgo }
In the ACS Virgo cluster survey Mei \etal~ (2007) provide
HST/ACS imaging for 100 early type  galaxies
and derives distances for 84 from Surface Brightness Fluctuations.
Five galaxies have d $\approx$ 23 Mpc and are members of the W$^\prime$ Cloud. 
For the remaining 79 galaxies the mean distance is 16.5$\pm$0.1 (random) $\pm$ 1.1 Mpc (systematic).
The rms distance scatter is 0.6$\pm$0.1 Mpc. 
The back-to-front depth of the cluster is 2.4$\pm$0.4 Mpc. The 
M87 (Cluster A) and M49 (Cluster B) subclusters lie at 
16.7$\pm$0.2 and 16.4$\pm$0.2 Mpc, respectively. 
Virgo's early-type galaxies appear to define a slightly triaxial distribution, with axis ratios of (1:0.7:0.5). 

Following this, we have the Next Generation Virgo Survey (Ferrarese \etal~ 2012). 
They anticipate distance errors of 0.13 mag from the NGVS data, corresponding to 1 Mpc. It is 
possible to study how galaxy properties correlate with the true $local$ environment determined from 3D positions.
They expect distances for more than 200 galaxies, including spatial segregation by galaxy type.

The M87 distance from an HST CMD is provided by 
Bird \etal~ (2010).
A deep (V,I) imaging dataset for the M87 with HST ACS resolved its brightest red-giant stars. 
As a byproduct, they obtained a preliminary measurement of the distance to M87 with the TRGB  method; 
the result was (m - M)$_0$ = 31.12$\pm$0.14 (d = 16.7$\pm$0.9 Mpc).

For TRGB and Cepheids
Mould \& Sakai (2008, 2009) find that when the tip of the red giant branch is used as the standard candle, 
the value of the Hubble constant is the same as when Cepheid stars are used. This finding is in agreement
with Tammann \& Reindl (this volume).
This  confirms the findings of the Hubble Space Telescope distance scale Key Project.
That finding is in disagreement with Tammann \& Reindl.

The Globular Cluster Luminosity Function is not elsewhere reviewed in this Symposium.
Villegas \etal~ (2010) studied
the GCLF turnover magnitude, $\mu_z$, for all the galaxies in the ACS Virgo and Fornax cluster surveys. They found 
$\mu_z$ = (23.51 $\pm$ 0.11) + (0.04 $\pm$ 0.01)M$_{z,gal}$, plus an offset of $\delta$(m - M) = 0.20 $\pm$ 0.04 mag 
for the galaxies in Fornax.

We can also ask if secondary distance indicators continue to be necessary if Cepheids can be found and measured in the 
Coma cluster. Samantha Hoffmann will present these in her PhD thesis talk at the AAS meeting at the end of the year.
The NGC 4921 HST/ACS team is Gregg, Cook, Macri, Mould, Stetson, Welch and Hoffmann. The observing bandpass is F350LP, 
which is practically ACS white light.

\section{Replacing Virgo with the ensemble of local supercluster distances}
Figure 2 shows the velocity field that allows Local Supercluster distances to be used as an ensemble to supplement
 the Virgo cluster distance in the traditional distance scale ladder. A similar velocity field has been calculated by Lavaux \& Tully (2010) also from the 2MRS survey. The corresponding distance catalog is compiled from the following sources.
\begin{itemize}
\item HST Cepheids
\item EDD (Tully)\footnote{http://edd.ifa.hawaii.edu/}
\item SDSS fundamental plane (George\footnote{http://physics.uq.edu.au/ap/cosmicflow/?page\_id=14}
)
\item Feldman, Watkins, \& Hudson (2010) catalog
\item Nearby SNIa distances
\item SBF distances (Tonry 2001)
\item SFI++ (Springob et al 2007)
\item NED database
\end{itemize}
One can, however, ask, does mass follow light?
Reconstructions are based on the assumption that it does, but
chinks in that argument exist in the separation of dark and luminous matter in the bullet cluster and in
 evidence for bulk flows not predicted by 2MRS (Magoulas 2012).

One may also ask about the uncertainties in this velocity field.
One answer is consistency with the Lavaux \etal~ (2010) velocity field which is also based on 2MRS.
Comparisons between Erdogdu and Lavaux show a lot of scatter.
Another answer is to recompute Erdogdu's velocities with monte carlo style perturbations to it. 
A third answer is to adopt some heuristic errors scaled to the density of the region of any particular
galaxy of interest. For example, in cluster density regions (e.g. Virgo) (where there tend to be few spirals anyway)
the uncertainty is as large as the cluster velocity dispersion, because the region is virialized. Indeed,
in such regions one is best off using the cluster distance for all the
galaxies within a projected Mpc or so. Then there are very low density regions, where computed velocities should
be near perfect (modulo the uncertainty in biassing). And there are intermediate density regions where the uncertainty would be somewhere in between.
Future improvements to the velocity field will come from
\begin{itemize}
\item More distances: e.g.
WALLABY/SkyMapper\footnote{http://db.ipmu.jp/seminar/sysimg/seminar/666.pdf}
and WNSHS/PanSTARRS\footnote{http://www.astron.nl/$\sim$jozsa/wnshs/}
\item Deeper redshift surveys: e.g. 
TAIPAN (UKST),\footnote{http://physics.uq.edu.au/ap/cosmicflow/?page\_id=14}
the ongoing 2MRS (Huchra \etal~ 2012)
and possibly future work at Apache Point Observatory.
\item Better velocity calculations beyond the linear approximation and employing higher resolution.
\end{itemize}

\section{Myths}
A long standing furphy is that there is no such thing as the Hubble Constant. Density variations make the expansion rate a function of position H($\theta,\phi$,z).
A recent misunderstanding is that the traditional distance ladder has been replaced by fitting CMB anisotropies and what they evolve into. The sound horizon is the new standard ruler.
A third fundamentalist (or extreme empiricist) view is that H$_0$ from CMB anisotropies is model dependent and not strictly relevant to measuring the Hubble Constant.
 
To these misconceptions, we respond, discard the myths. Embrace the orthodoxy: the standard model of cosmology.

In this view of the distance scale, which is widely shared,
H$_0$ is the first parameter in a model of an evolving expansion rate.  In real experiments there is a covariance between it and other parameters.
There are a number of observational constraints on this parameter. They include, but are not confined to, measurements at z $>$ 0.
Variations of density from the mean can be mapped and corrections to a global expansion rate made to local measurements.

A suitable conclusion to a digression on myths is the steady state universe. Its fit to supernova data is shown in Figure 3. To plot the model, I have set $\Omega_3$ =0 in the polynomial Friedmann equation of Mould (2011) and $\Omega_0$ = $\Omega_M$. This corresponds to a constant matter density which does not decrease as (1+z)$^3$. The result is a curvature dominated universe. In resurrecting the steady state universe, one discards conservation of mass. In accepting the standard model of cosmology, one accepts a 70 order of magnitude discrepancy between predicted and observed vacuum energy (Martin 2012). However, even if one were to make a choice between these two evils, it is clear from Figure 3 that the steady state universe does not even fit the data for z $<$ 1, let alone the microwave background.

\section{New motivations for an accurate distance scale}
The WMAP model has 6 free parameters, one of which is the number of relativistic species in the earlier 
thermal history of the Universe. Steigman (2012) summarizes the evidence that there are 4 `neutrinos'.
Other evidence is presented by
Suyu \etal~ (2012) and Freedman \etal~ (2012). If these results are to be raised to the mandatory (Higgs) 5$\sigma$ significance, the Hubble Constant will need to be measured to 1\%.									 
Allan Sandage, to whom this symposium is dedicated, famously described cosmology as the search for two numbers (Sandage 1961).
So, he did not subscribe to any of the myths listed in the previous section. Now, there are 6 numbers.
It is a safe prediction that there will be more parameters required to compose a theory of physical cosmology
and also that they will steadily become better determined through observations.

\section{Summary}
The Virgo cluster once had an iconic status in the extragalactic distance scale ladder.
In its place we now have an ensemble of local supercluster distances, connected by a velocity field based
on redshift survey reconstructions.
The ladder is now just one of a number of experiments aimed at measuring H$_0$, which is one of 6 parameters in the current standard model of cosmology.
Other experiments are the CMB and Baryon Acoustic Oscillations. The ladder itself may be bifurcating
 into the Cepheid-SNIa ladder and a megamaser-BAO ladder. The former may be the luminosity distance scale.
The latter is an angular diameter distance scale.
Parameters in the standard model have covariance (degeneracies).
This leads to new reasons to constrain H$_0$ to 1\% (e.g. \# neutrinos).

\acknowledgements
I would like to thank Roger Blandford for hosting a workshop at the Kavli Institute (KIPAC) at Stanford in February,
where a number of these ideas were developed. I'd also like to recall John Huchra's
contributions to the distance scale up to his untimely death in 2010. The study of the dark universe
is supported by the Australian Research Council through CAASTRO\footnote{www.caastro.org}.
The Centre for All-sky Astrophysics is an Australian Research Council Centre
of Excellence, funded by grant CE11001020.

\section*{References}

\noindent                       
Aaronson, M. \& Mould, J. 1983, ApJ, 265, 1\\      
Bird, S., Harris, W., Blakeslee, J., \& Flynn, C. 2010, A\&A, 524, 71\\	
Binggeli, B., Sandage, A. \& Tammann, G. 1985, AJ, 90, 1681\\
Bohringer, H.,  Briel, U.,  Schwarz, R.,  Voges, W.,  Hartner, G.,  \& Trumper, J.1994, Nature, 368, 828\\	
Catinella, B. \etal~ 2012, MNRAS, 420, 1959\\      
Erdogdu, P. \etal~	 2006, MNRAS, 373, 45\\	
Feldman, H., Watkins, R., \& Hudson, M., 2010, MNRAS, 407, 2328\\
Ferrarese, L. \etal~ 2012 ApJS, 200,4\\ 
Freedman, W. \etal~ astro-ph 1208.3281.\\          
Huchra, J. P., Macri, L., Masters, K. \etal~ 2012, ApJS, 199, 26\\	      
Lavaux, G., Tully, R.B., Mohayaee, R. \& Colombi, S., 2010, ApJ, 709, 483\\
Magoulas, C. 2012, Ph.D. thesis, University of Melbourne\\
Martin, J. 2012, astro-ph 1205.3365\\ 
Mei, S. \etal~ 2007, ApJ, 655, 14\\
Mould, J. 2011, PASP, 123, 1030\\
Mould, J., Aaronson, M. \& Huchra, J. 1980, ApJ, 238, 458	\\
Mould, J. \& Sakai, S. 2009, ApJ, 697, 996\\         
Mould, J. \& Sakai, S. 2009, ApJ, 694, 1331\\
Mould, J. \& Sakai, S. 2008, ApJ, 686, L75\\	
Sandage, A. 1961, ApJ, 133, 355\\
Springob, C. \etal~ 2007, ApJS, 172, 599\\
Steigman, G. 2012,  astro-ph 1208.0032\\
Suyu, S. \etal~ astro-ph 1202.4459\\
Tammann, G. 1999, IAU Symposium, 183, 31\\
Tonini, C. \etal~ 2012, in preparation\\	
Tonry, J., Dressler, A., Blakeslee, J., Ajhar, E., Fletcher, A., Luppino, G., Metzger, M., \& Moore, C. 2001, ApJ, 546, 681\\
Villegas, D. \etal~ 2010, ApJ, 717, 603\\ 


\begin{figure}[b]
\begin{center}
 \includegraphics[width=3.4in,angle=-0]{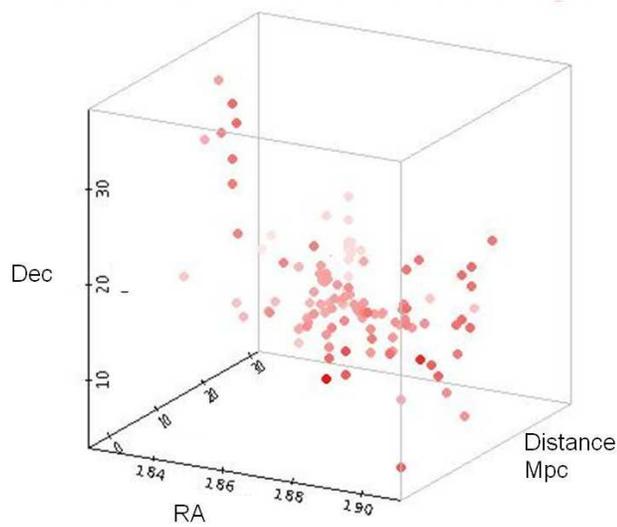} 
 \caption{The structure of the Virgo cluster from the NED 1D database.}
\end{center}
\end{figure}

\begin{figure}[b]
\begin{center}
 \includegraphics[width=6.4in,angle=-90]{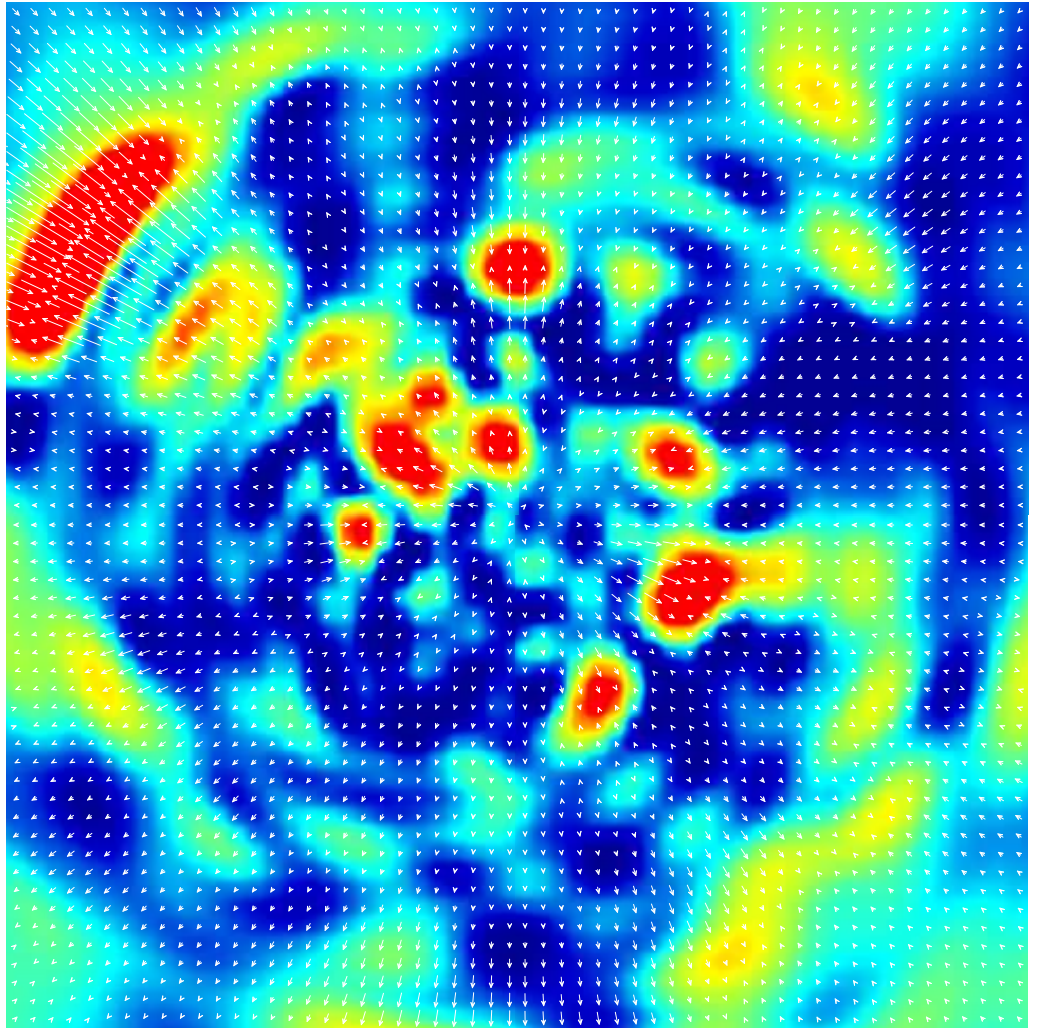} 
 \caption{The density field (coloured) in the supergalactic plane and the velocity field within 150 Mpc from 2MRS,
Erdogdu \etal~ (2006). We at the origin are at the centre.
The largest structure shown is the Shapley Supercluster.}
\end{center}
\end{figure}
\pagebreak

\begin{figure}[b]
\begin{center}
 \includegraphics[width=4.4in,angle=-90]{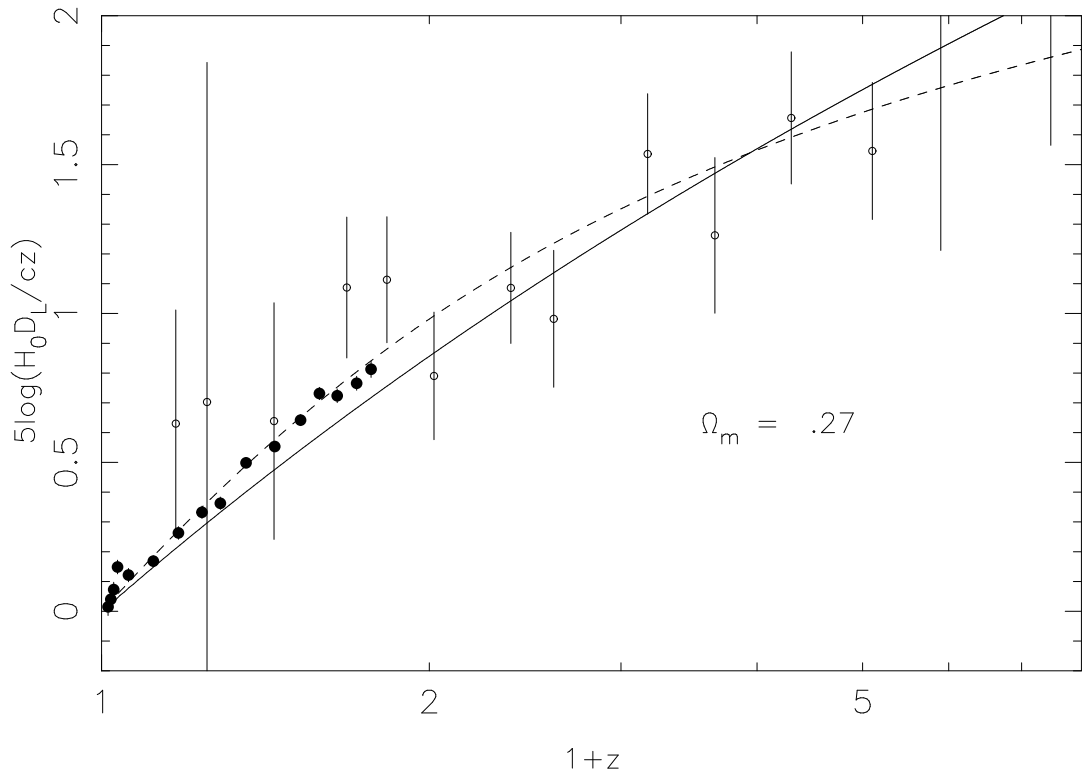} 
 \caption{The redshift distance relation for a steady state
universe with $\Omega_M$ = 0.27. This is a curvature dominated
Friedmann equation. Supernova and GRB data are the points.
For comparison, the standard model of cosmology with dark energy
and zero curvature is shown as a dashed line. }
\end{center}
\end{figure}



\begin{discussion}

\discuss{Michael Feast}{Your F350LP observations are going to have problems with calibration and extinction.}
\discuss{Jeremy Mould}{We'll work on these in the next 6 months and also work on photometric simulation. Observations at V \& I have also been made.}
\discuss{Giuseppe Bono}{Your Coma magnitudes seem to differ from what I'd expect.}
\discuss{Jeremy Mould}{Our preliminary Virgo--Coma modulus within the current uncertainties is 
consistent with the ratio of Coma's redshift and Virgo's flow corrected redshift.
When we've taken the steps I just alluded to, the uncertainty in NGC 4921's distance will drop
to perhaps 5\% (I'm guessing) and then we'll see.}

\end{discussion}
\end{document}